\documentclass[twocolumn,showpacs,preprintnumbers,amsmath,amssymb,floatfix]{revtex4}

\pdfoutput=1

\usepackage{graphicx}
\usepackage{dcolumn}
\usepackage{bm}
\usepackage[latin1]{inputenc}
\usepackage{float}
\usepackage{verbatim}

\begin{document}

\title{Metastable states influence on the magnetic behavior of the triangular lattice:
       Application to the spin-chain compound Ca$_3$Co$_2$O$_6$}

\author{R. Soto,$^{1}$ G. Mart\'{\i}nez,$^{1,2}$ M. N. Baibich,$^{2}$
        J. M. Florez,$^{1}$ and P. Vargas$^{1,3,}$}

\email{vargas.patricio@gmail.com}

\affiliation{
   $^{1}$Departamento de F\'{\i}sica, Universidad T\'ecnica Federico Santa Mar\'{\i}a,
   P.O. Box 110-V, Valpara\'{\i}so, Chile \\
   $^{2}$Instituto de F\'{\i}sica, Universidade Federal do Rio Grande do Sul,
   91501-970 Porto Alegre-RS, Brasil\\
   $^{3}$Max-Planck Institute for Solid State Research, Heisenbergstrasse 1,
   D-70569, Stuttgart, Germany}

\date{\today}

\begin{abstract}
It is known that the spin-chain compound Ca$_3$Co$_2$O$_6$ exhibits very
interesting plateaus in the magnetization as a function of the magnetic
field at low temperatures. The origin of them is still controversial. In
this paper we study the thermal behavior of this compound with a single-flip
Monte Carlo simulation on a triangular lattice and demonstrate the decisive
influence of metastable states in the splitting of the ferrimagnetic 1/3
plateau below 10~K. We consider the [Co$_2$O$_6$]$_n$ chains as giant magnetic
moments described by large Ising spins on planar  clusters with open boundary
conditions. With this simple frozen-moment model we obtain stepped magnetization
curves which agree quite well with the experimental results for different sweeping
rates. We describe particularly the out-of-equilibrium states that split the
low-temperature 1/3 plateau into three steps. They relax thermally to the 1/3
plateau, which has long-range order at the equilibrium. Such states are further
analyzed with snapshots unveiling a domain-wall structure that is responsible
for the observed behavior of the 1/3 plateau. A comparison is also given of the
exact results in small triangular clusters with our Monte Carlo results, providing
further support for our thermal description of this  compound.
\end{abstract}


\maketitle

\section{Introduction}

Low-dimensional interacting spins always reveal very interesting
magnetic properties, as well as new electronic transport phenomena
\cite{Oshikawa,Cabra,Vekua,Hamad}. Along this line, systems with
geometric frustration have long attracted our attention because the
ground-state properties, like degeneracy, are usually responsible
for peculiar behaviors at low temperatures. Included  are exotic
magnetization dependences with external fields in hysteresis
curves, for example. One group of compounds that exhibits such
properties is the family CsCoX$_3$, where X stands for Cl or
Br \cite{Collins}. These materials are unidimensional Ising-like
magnetic systems. The spin chains are arranged on a triangular
lattice, while the intra- and interchain exchange couplings are
both antiferromagnetic.

\begin{figure}[h]
\begin{center}
\includegraphics[scale=0.38,angle=0]{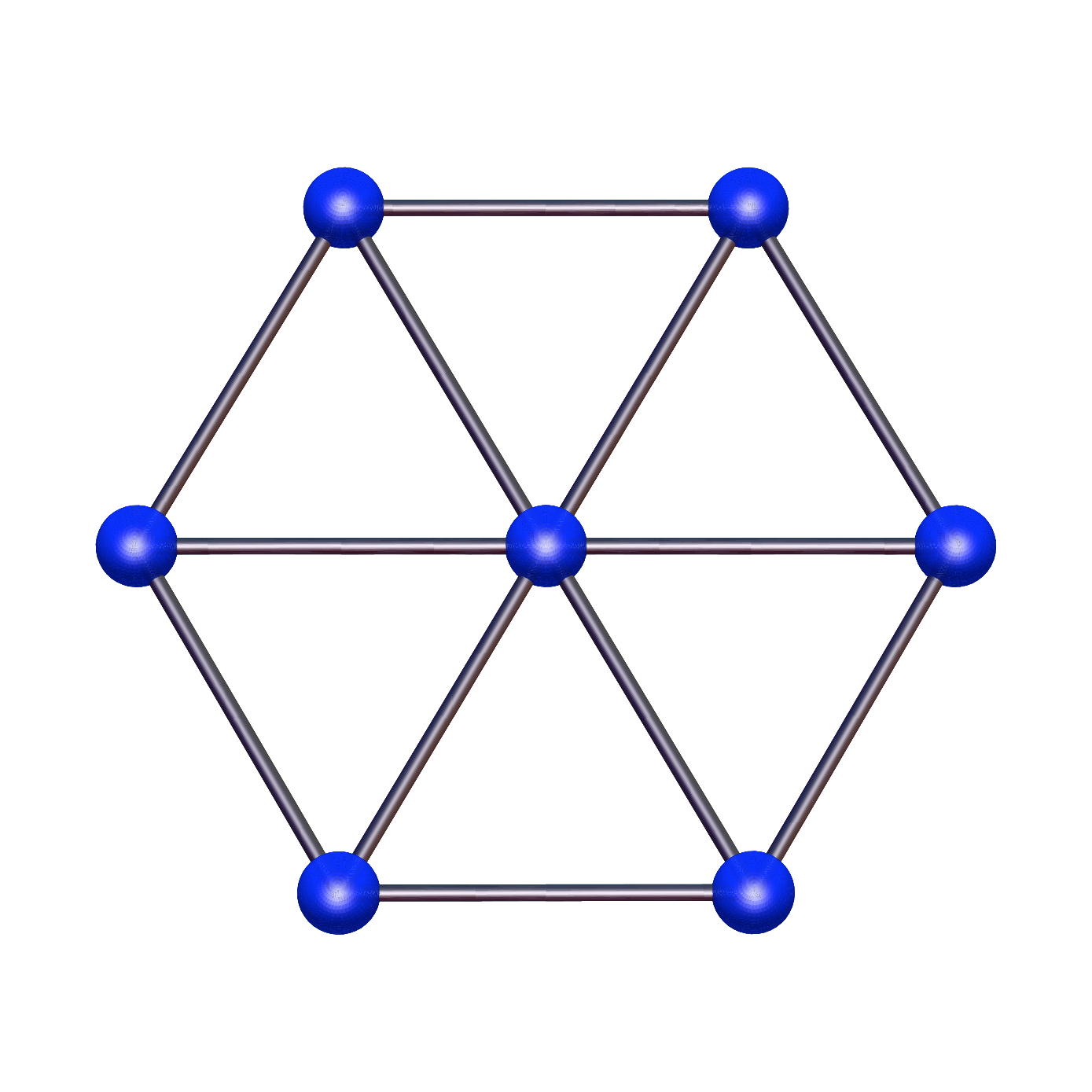}
\end{center}
\caption{Hexagonal lattice of Co chains projected onto the $ab$-plane
(blue points) of the spin-chain compound Ca$_3$Co$_2$O$_6$. Each chain is
ordered ferromagnetically and is coupled antiferromagnetically to its neighbors.}
\label{hexagono}
\end{figure}

Another fascinating  spin-chain family of compounds with triangular
arrangements has the formula A'$_{3}$ABO$_6$ (where A' may be Ca or
Sr, whereas A and B are transition metals). One particular system that
caught our attention is Ca$_3$Co$_2$O$_6$ \cite{Fjellvag,Aasland,Kageyama,
MaignanEPJB152000,martinez,Raquet,hardy2003,vidya,hardy2,whanboo,QTMMaignan,
fresard,eyert,MaignanPRB702004,flauhaut,hardy3,villesuzanne,takubo,yamada},
which has a rombohedral structure composed of large chains of [Co$_2$O$_6$]$_n$
along the $c$-axis of a corresponding hexagonal lattice. The Ca atoms are
located among those chains. The chains are made of alternating face-sharing
CoO$_6$ trigonal prisms and CoO$_6$ octahedra. Each Co chain is thus
surrounded by six equally spaced Co chains forming an hexagonal lattice
in the $ab$-plane (see Figure \ref{hexagono}).

Generally, the long Co-Co interchain distance (5.3~\AA), {in Ca$_3$Co$_2$O$_6$}, compared to
the short Co-Co intrachain distance (2.6~\AA), ensures the hierarchy
of the magnetic exchange energies, $J_{\rm intra}\gg J_{\rm inter}$,
in modulus. This fact, together with a strong spin-orbit coupling
\cite{Wu,Wu2}, provide a high uniaxial anisotropy along the $c$-axis.
Neutron diffraction studies \cite{Aasland}, among others, have
revealed a ferromagnetic intrachain  interaction ($J_{\rm intra}<0$),
while an antiferromagnetic interchain interaction ($J_{\rm inter}>0$)
was observed in the $ab$-plane. Consequently, it is very plausible to
describe the magnetism of this compound by considering an Ising triangular
lattice of large magnetic frozen-moments linked to the chains,
each one playing the role of a giant spin, pointing either up or down.

\newpage

One of the interesting features observed in Ca$_3$Co$_2$O$_6$ is
the appearance of plateaus in the magnetization curve $M$ vs $H$, below
$T_C=25$~K, when the magnetic field is applied along the chains ($c$-axis).
For temperatures in the range, $10\, {\rm K} \leq T \leq T_C$, it is observed
a rapidly increase of $M$ with $H$. Close to zero field, the magnetization
reaches a plateau at the value of $M_s/3$ ($M_s$, the saturation magnetization
is $\sim 4.8\,\mu_{\rm B}$/f.u.). This value remains constant up to
$H_c \approx 3.6$~T, where the magnetization springs up to its saturation value
$M_s$. Otherwise, when $T\leq 10$~K, called the ferrimagnetic region, the observed
plateau at $M_s/3$, splits into three small steps, changing at fields $H=1.2$~T,
$2.4$~T, $3.6$~T, respectively. Much effort has been done to understand such behavior,
and it is still a controversial matter. Some papers \cite{QTMMaignan,hardy3,yamada}
roused the question that this behavior might be related to experiments involving
single-molecule magnets (SMM) \cite{smm1,smm2}, where a Quantum Tunneling of the
Magnetization (QTM) is realized \cite{gunther}. We restrain such statement, at least
for temperatures above 4K, as we demonstrate here and leave the question of QTM effects to
be searched below 2K, according to the experimental situation.

We follow instead the line of reasoning that metastable states, or peculiar mozaic
configurations, might play a role in the splitting of the magnetization of this
frustrated material, as first suggested by Maignan {\it et al.} \cite{MaignanEPJB152000}.
Therefore, we started a simulation along the lines given by Kudasov \cite{kudasov},
who predicted an opening of the 1/3 plateau using Monte Carlo analysis at $T=0$. More
recently, Yao {\it et al.} \cite{YaoPRB732006,YaoPRB742006}, studied this compound
with numerical Monte Carlo simulations on finite clusters. The latter works describe
the system with an Ising Hamiltonian on a triangular lattice, using periodic boundary
conditions. Despite their results reproduce the steps width, $\Delta H\sim 1.2$ T, in
the magnetization curves, they do not find a clear-cut ({\it equilibrium}) configuration
that explains the splitting of the 1/3 plateau. In this work we report on a similar
study using the same model, but considering instead free boundary conditions on an
anisotropic system, which is added for completion with the intrinsic dipolar term at
the end. We apply non-equilibrium techniques on this system. We detect in this form,
by sweeping the field at different rates, the presence of metastable domains-walls that
perfectly explains the observed trends of the splitting of the 1/3 plateau at low
temperatures, and verify that the latter has long-range order at the equilibrium.
To be consistent with the previous existing models, we use the same set of parameters
for the Hamiltonian \cite{kudasov,YaoPRB732006,YaoPRB742006}, namely, an interchain
antiferromagnetic interaction, $J=J_{\rm inter}=2.25$ $\mu$eV, and a large Ising magnetic
moment, $S=32$. Disorder or dispersive effects in the exchange coupling constant were not
required to obtain the splitting of the 1/3 plateau, as it will become clear in the following.
Therefore, our model and methods are different from Yao {\it et al.} \cite{YaoPRB732006,YaoPRB742006}, and
we concentrate our efforts mainly to describe the effects of the sweeping rate on the
observed metastable states, that generate the sub-steps in the magnetization.

\section{Monte-Carlo Simulations in 2D}

The spin-chain compound Ca$_3$Co$_2$O$_6$ is formed by close-packed one-dimensional
chains along the $c$-axis, the chains span a triangular lattice over the $ab$-plane.
Due to such geometric configuration and the hierarchy of the exchange interactions,
the system can be modeled as a two-dimensional (2D) antiferromagnetic Ising model
with nearest-neighbor exchange interactions \cite{kudasov,YaoPRB732006,YaoPRB742006}.
The magnetic moments are representative of the [Co$_2$O$_6$]$_n$ chains, which are
always perpendicular to the $ab$-plane. The choice of this frozen-moment model is
a consequence of a time-scale assumption. We believe that the characteristic time
involved in reversing the individual spins of each chain is much shorter than the
rearrangement of the triangular lattice by changing the magnetic field. This hypothesis
is based on previous numerical results \cite{nanowires} on nano-wires, as well as on
the large number of sites that are involved in changing the whole triangular
lattice as compared to those on a single spin-chain.

We thus performed Monte Carlo simulations using a triangular Ising model, where the magnetic
moments are frozen, pointing either up or down, and coupled antiferromagnetically ($J>0$)
to their nearest neighbors. In this picture, the Hamiltonian $\cal H$ of the system is

\begin{equation}\label{Ising Hamiltonian}
    {\cal H} =  J \sum_{\langle i,j\rangle} S^z_i  S^z_j-
    g\mu_{\rm B}  \sum_{i} {\bf S}_i \!\cdot\! {\bf H} \,,
\end{equation}
where ${\bf S}_i$ is the spin of each magnetic moment, $\bf H$ is the external magnetic
field along the $+c$-axis, $g = 2$ is the electronic Land\'e factor, $\mu_{\rm B}$ the Bohr
magneton and $\langle i,j\rangle$ indicates summation over all pairs of nearest-neighbors
on a triangular lattice. We used a single-flip Monte Carlo method together with the standard
Metropolis algorithm to reject/accept flips in the magnetization with a Boltzmann factor
\cite{Binder}. Magnetization measurement at any given field was done using 1,000 averages
over a fixed number of {\it accepted} Monte-Carlo (MC) flips.
We used different 2D finite cluster configurations, including super-hexagons, octagons,
dodecagons and other more circularly-shaped geometries to represent the samples.
The magnetization curves found in all these finite clusters were almost the same, the main
difference was the number of MC flips needed to reach the plateaus for different sizes.
Due to space limitations and visualization aspects, a selection of the results for one cluster
size with hexagonal geometry was chosen. Extra material showing the evolution of magnetic
configurations with the field can be found at the website by the interested
reader [supplementary material].

\begin{figure}[ht]
\begin{center}
\includegraphics[scale=0.6,angle=0]{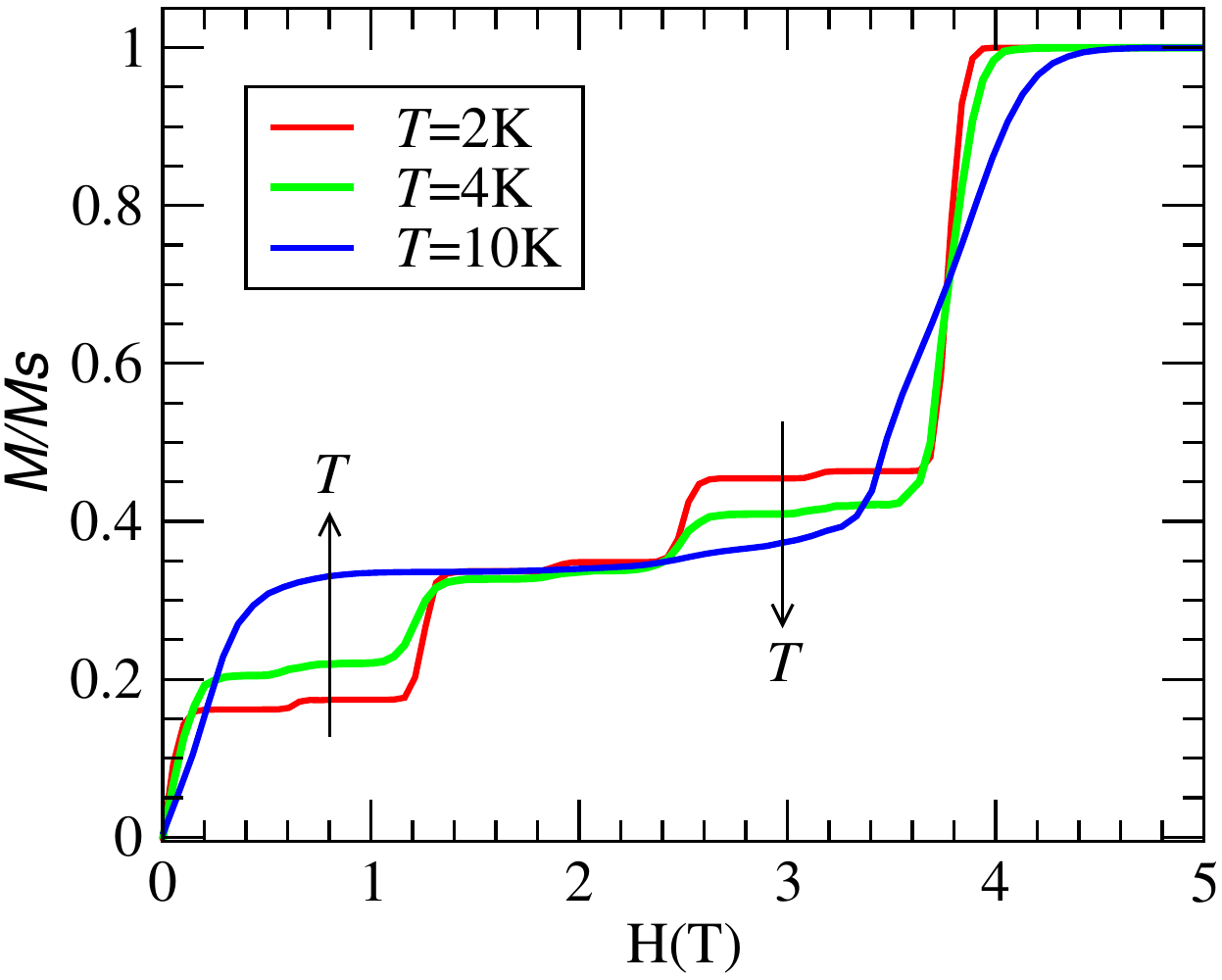}
\end{center}
\caption{MC results for the magnetization as a function of the applied field
for three different temperatures in the ferrimagnetic phase, $T\leq 10$~K,
and $N=2431$ sites. The splitting of the 1/3 plateau into three successive
plateaus is observed below 10 K. They extend up to fields $H\approx 1.2$~T,
2.4~T and 3.6~T, respectively. We see that they merge into a unique plateau
at $M\approx M_s/3$ as $T$ increases (see arrows).}
\label{Tvar}
\end{figure}

\section{Magnetization Results}

In this section we discuss our magnetization results by using Monte Carlo
calculations on finite clusters with open boundary conditions, always
increasing the field, starting at $ H=0$ T. The initial magnetic state was obtained
by thermalizing a random disordered configuration, performing $10^6$ MC flips in
a {\it training} period. Afterwards we initiated the increase of the field
at a constant sweep rate.
We calculated clusters with hexagonal geometries for $N= 1387$, $2431$,
$3169$, $4921$ and $6931$ sites. Some other larger clusters were partially
calculated, but we verified that the agreement with the experimental results
did not improve with size. As the system size increased the plateaus structure
remained unaltered, only small perturbations (sub-steps) on each of the three main
plateaus tend to disappear.

In Figure \ref{Tvar} we readily observe the formation of three
well defined plateaus below $T=10$~K. Important to see is that
they merge into one, at $M_s/3$ as $T$ increases, indicating a
close resemblance with the experimental situation
(see e.g. Fig.\ 3 in \cite{MaignanPRB702004}).
The presence of extra tiny sub-steps is an unavoidable boundary effect
due to the finiteness of the cluster used (in this case, $N=2431$ sites).
They are created by border isolated atoms whose contribution is
strongly suppressed for larger cluster sizes, therefore,
irrelevant for our argument. Another important feature to note
in the results of Figure \ref{Tvar}, shared by experiments, is the
steeper rise of the magnetization for lower temperatures at the end
of the 1/3 plateau ($H_c\approx 3.6$~T).

\begin{figure}[ht]
\begin{center}
\includegraphics[scale=0.6,angle=0]{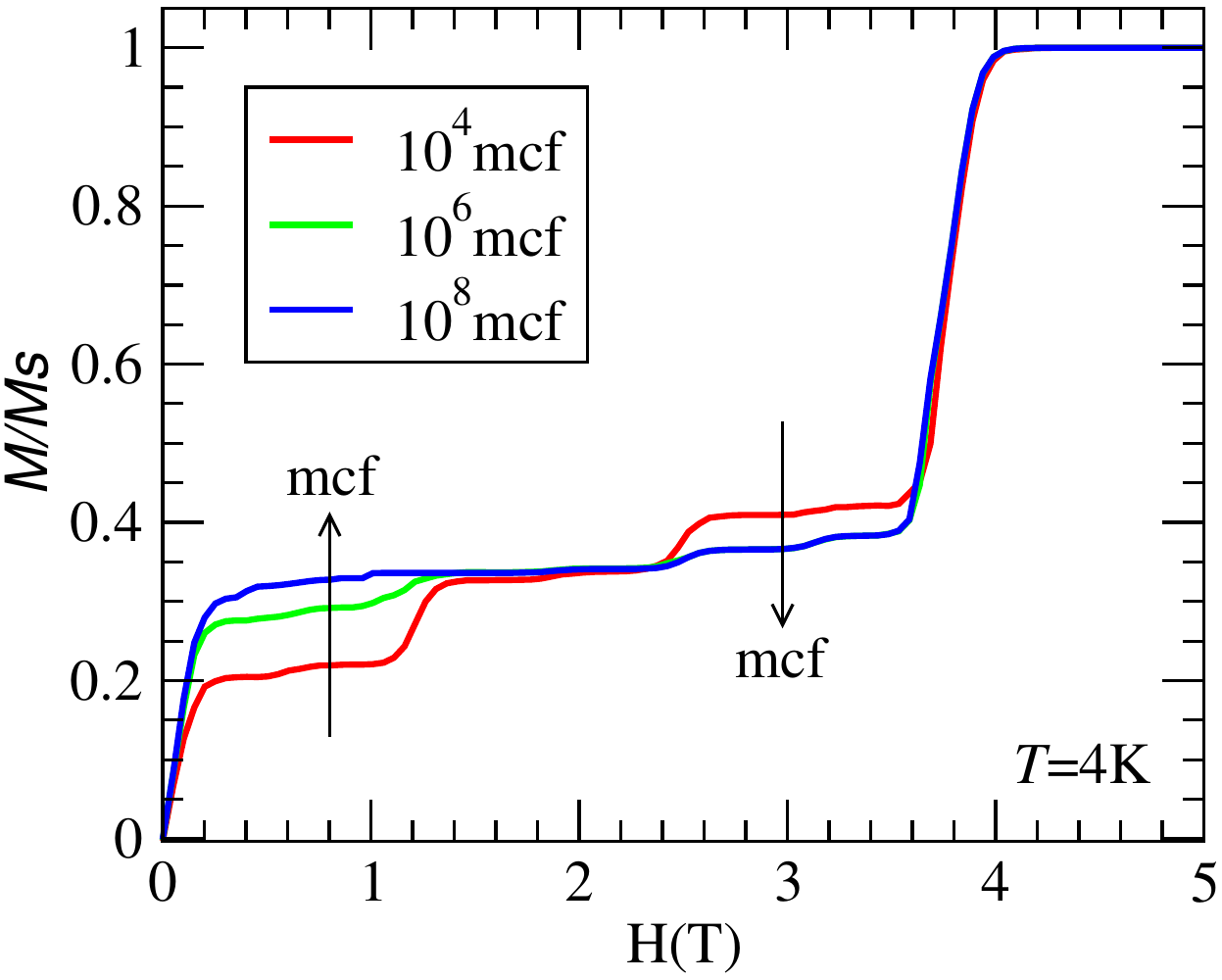}
\end{center}
\caption{MC results for the magnetization as a function of the applied
field for three different sweep rates, at $T=4$ K, and $N=2431$ sites.
This is accomplished by letting more or less accepted MC flips (mcf) per
magnetic field. Notice the tendency of merging into the plateau $M_s/3$ as
the number mcf increases (see arrows), indicating a time decay towards
the 1/3 state. This behavior will be further analyzed.}
\label{mcfvar}
\end{figure}

Now, as a way to check how stable are such plateaus, we have performed MC
analyses with an increasing number of steps per magnetic field. In fact,
as the MC simulation is done with the accepted configurations, what really
counts is MC flips (mcf) and not MC steps (mcs). We clearly see, in Figure
\ref{mcfvar}, the tendency to merge the three plateaus into one as a function
of the number of mcf. This certainly must be related to the experimental
sweeping rate \cite{MaignanPRB702004}. This connexion allows us to say that the
three plateaus are metastable states and are, therefore, a dynamical effect;
for this very reason the experimental magnetization curves are strongly
dependent on the sweeping rate \cite{MaignanPRB702004}.

\begin{figure}[ht]
\begin{center}
\includegraphics[scale=0.6,angle=0]{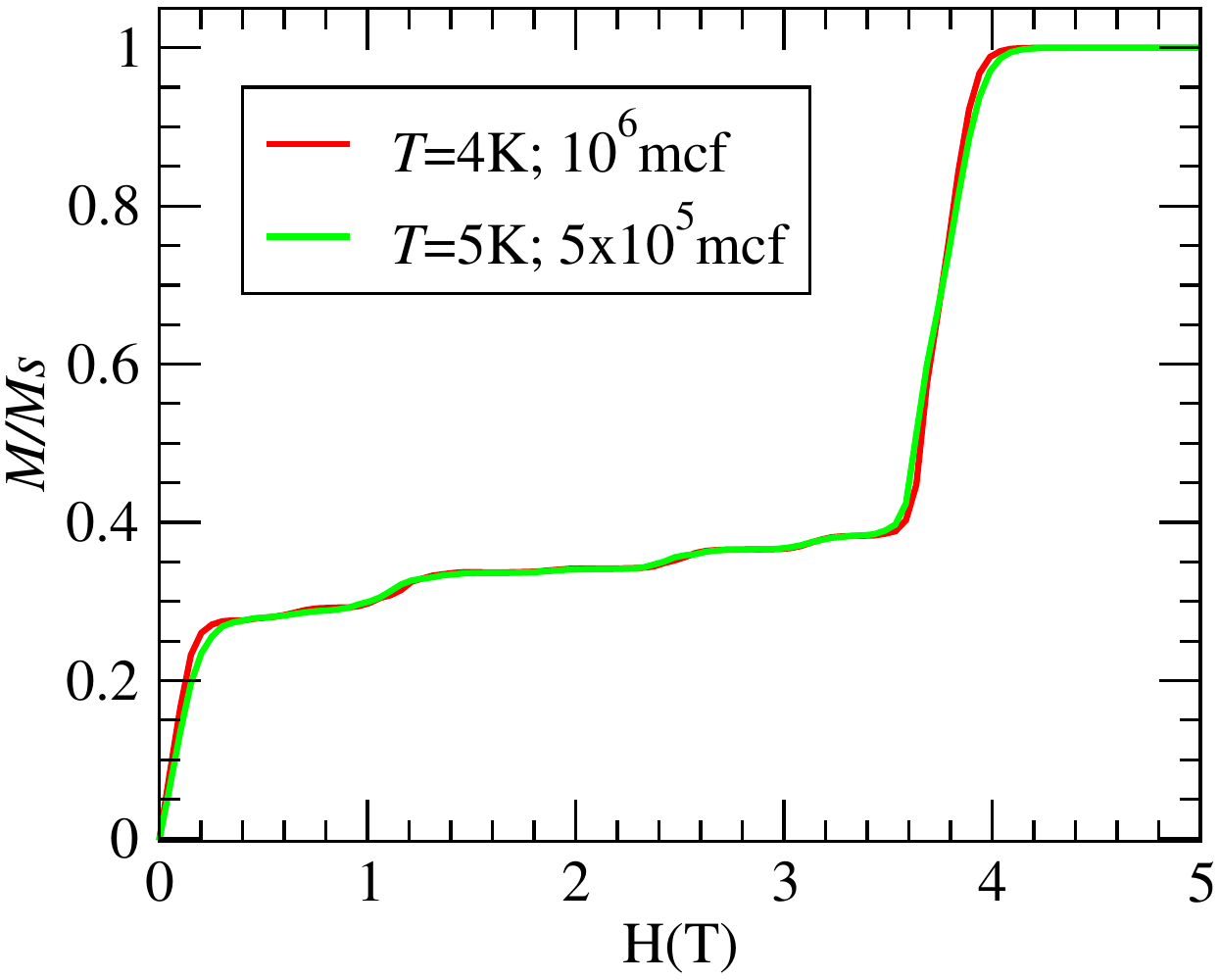}
\end{center}
\caption{Magnetization versus applied field for two different temperatures
and different sweep rates. We see that they merge into very similar curves.
This overlap indicates that they correspond to the same statistical configuration,
one which realizes similar magnetic moments. Such {\em scaling} is an extra
indication of metastable configurations. Differences are seen at both steep rising
regions, close to zero and at $H_c\approx 3.6$~T, fact which is also seen in
the experiments  \cite{MaignanPRB702004}.}
\label{T4K_5K}
\end{figure}

This merging effect can be further explained in the graph of Figure \ref{T4K_5K},
where we have produced an almost perfect superposition between magnetization
curves for different temperatures and, at the same time, different mcf. In such
a case, we tried to draw a parallel between the mcf in our simulation and the
sweeping rate in the experimental results, where this behavior is observed.
It is obvious from this result the strong interplay between the effects of time
and temperature (see e.g. Fig.\ 4 in \cite{MaignanPRB702004}), a feature that was
previously used as a fingerprint of QTM. Although mcf are {\it not} related to
lab time measurement, we can attribute them to some sweeping rate by comparing
the formation of the plateaus. Hence, the three sub-plateaus are made of metastable
states, which develop in time to the 1/3 plateau, as in Figure \ref{mcfvar}.

\begin{figure}[t]
\begin{center}
\includegraphics[scale=0.6,angle=0]{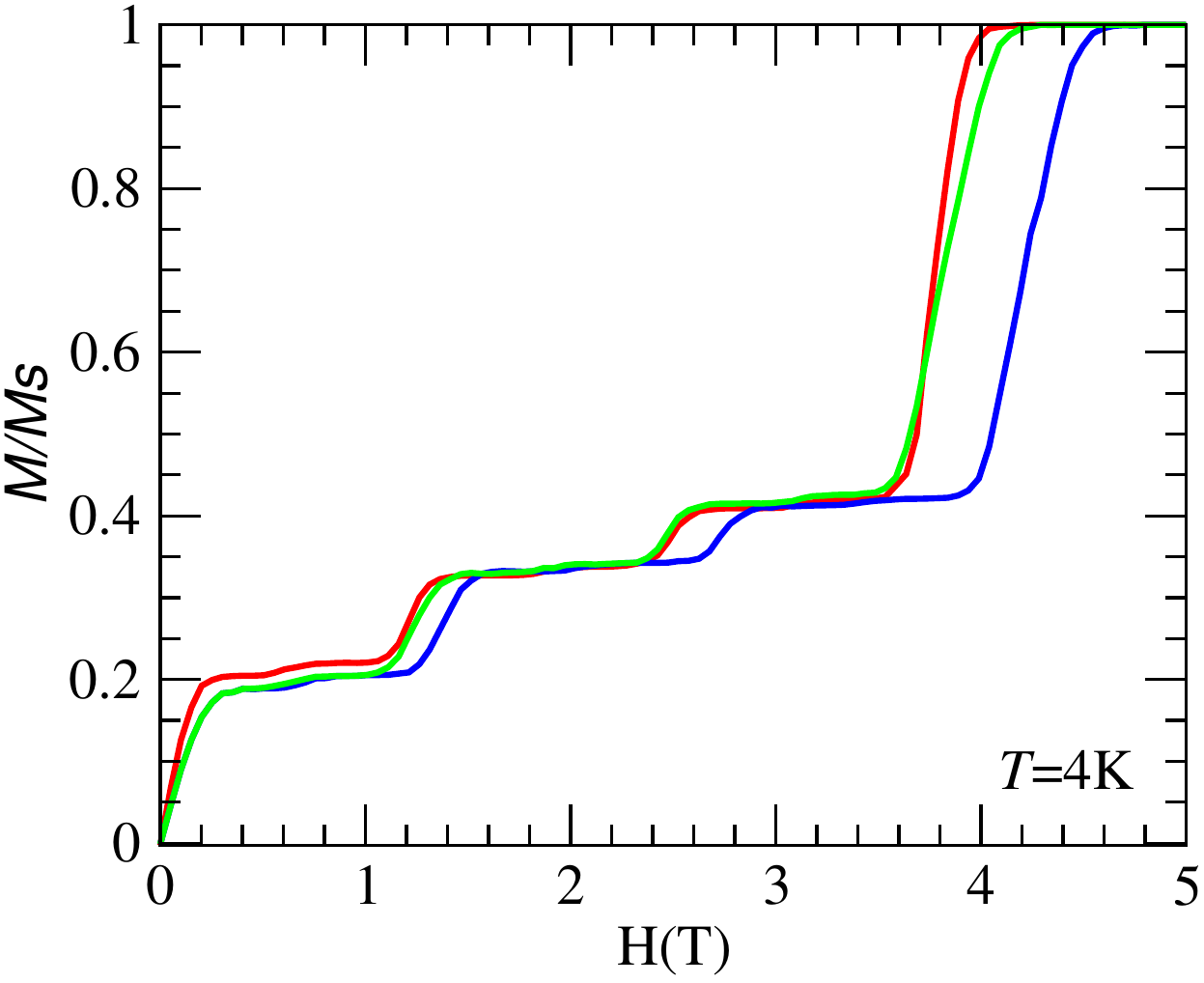}
\end{center}
\caption{Magnetization versus applied field considering dipolar
interactions, at $T=4$ K, and $N=2431$ sites. The red curve is for
the original $J$ value used in Figure \ref{mcfvar},
without dipolar interaction. The blue curve is for the same $J$,
but now with dipolar interaction, and the green curve
is for a reduced exchange,  90\% of the original value, and the
same dipolar interaction of the blue curve. The scaling of the
red and the green curves is not perfect, but very close.}
\label{dipolar}
\end{figure}

Dipolar interactions were also considered, as an intrinsic effect always present
in nanowires and other low-dimensional systems, but its overall effect here was
merely to induce an effective antiferromagnetic field that retarded the field at
which the steps occur. The dipolar term ${\cal H}_{\rm dip}$ that was added to
equation (\ref{Ising Hamiltonian}) is given by:
\begin{align}\label{dipolar Hamiltonian}
{\cal H}_{\rm dip}  = \frac{1}{2}\sum_{i,j}\frac{{\bf m}_i \!\cdot\! {\bf m}_j-3(
{\bf m}_i \!\cdot\!\hat{\bf r}_{ij})({\bf m}_j\!\cdot\!\hat{\bf r}_{ij})}{r_{ij}^{\,3}}
\end{align}
where the summation is over all sites, and ${\bf m}_i=\pm\mu_{\rm B} {\bf S}_i$.
We see such effect in Figure \ref{dipolar}, with the dipolar interaction playing
the role of an effective antiferromagnetic field, $H\rightarrow H+H_{\rm eff}$
added to equation (\ref{Ising Hamiltonian}), that moves the magnetization curves to
the right. We noticed that adjusting the exchange constant to $0.9J$, we can almost reproduce
the results obtained from our previous 2D Ising model, without dipolar interaction
(see Figure \ref{dipolar}). For this reason, we think, the magnetization curves are
well described using only the exchange and Zeeman terms.

\section{Snapshots and Domain-Walls}

The physics behind these results can be  visualized  by using raw
snapshots of selected configurations. In the next  pictures we show
different snapshots of the magnetization curves depicted with the red
line of Figure~\ref{mcfvar}. They were taken at three different magnetic
fields: $H=0.81$~T, $H=2.0$~T, and $H=2.98$~T, representative of each of
the three sub-steps, into which the 1/3 plateau is splitted at low
temperatures ($T<10$~K).

\begin{figure}[b]
\begin{center}
\includegraphics[scale=0.30,angle=0]{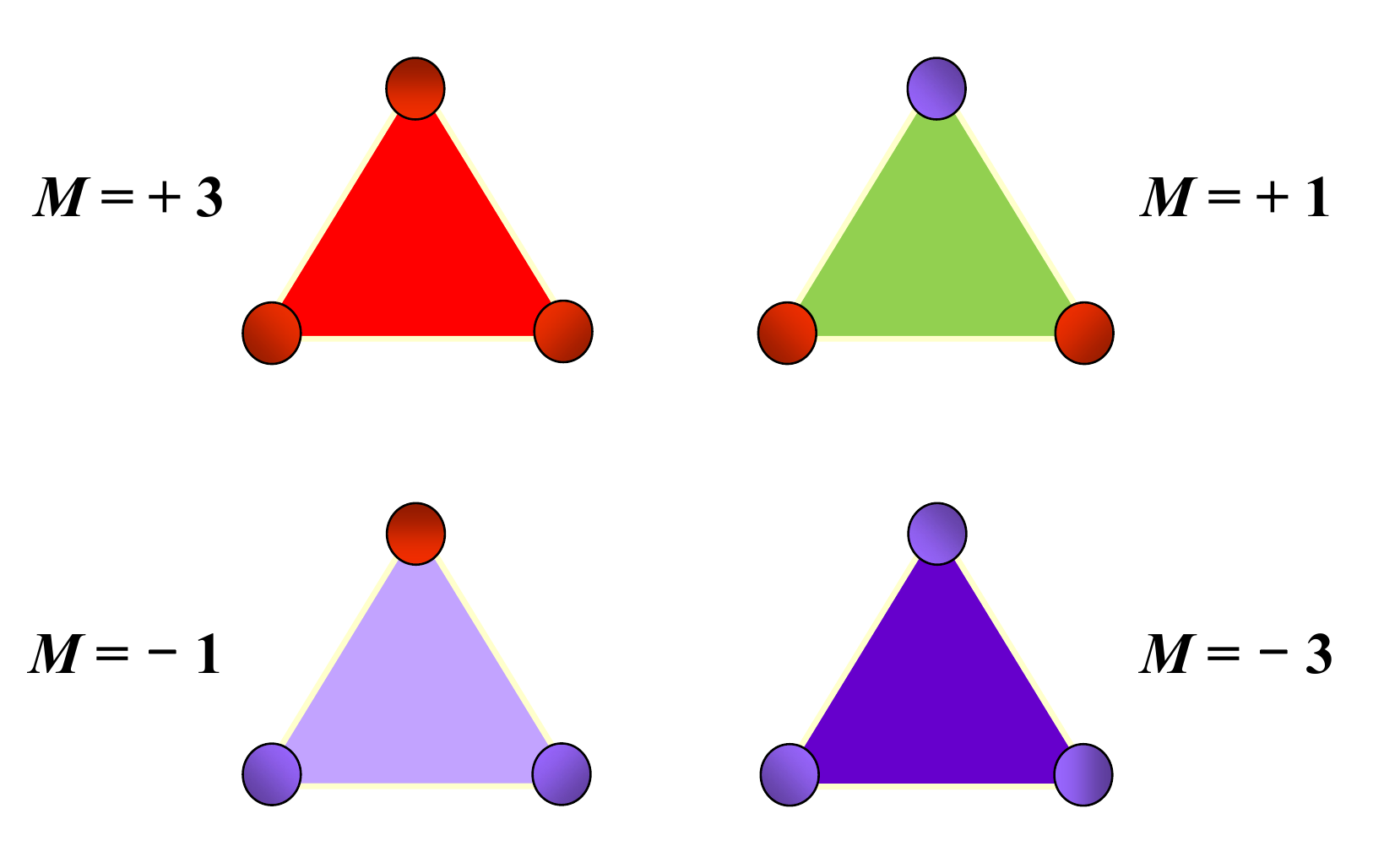}
\end{center}
\caption{The magnetic moment of each triangle depends on the
orientation of each ferromagnetic chain. The four possibilities
are schematized by different colors, but $M=-3$ is seldom seen
for positive magnetic fields in the following snapshots.}
\label{apices}
\end{figure}

Our color convention to analyze the snapshots of the subsequent examples
is depicted in Figure \ref{apices}. Following \cite{MaignanEPJB152000},
we define four triangles and color them differently depending on
the magnetization that results by summing the three spins at their
vertices:  see Figure \ref{apices}.

\begin{figure}[t]
\begin{center}
\includegraphics[scale=0.29,angle=0]{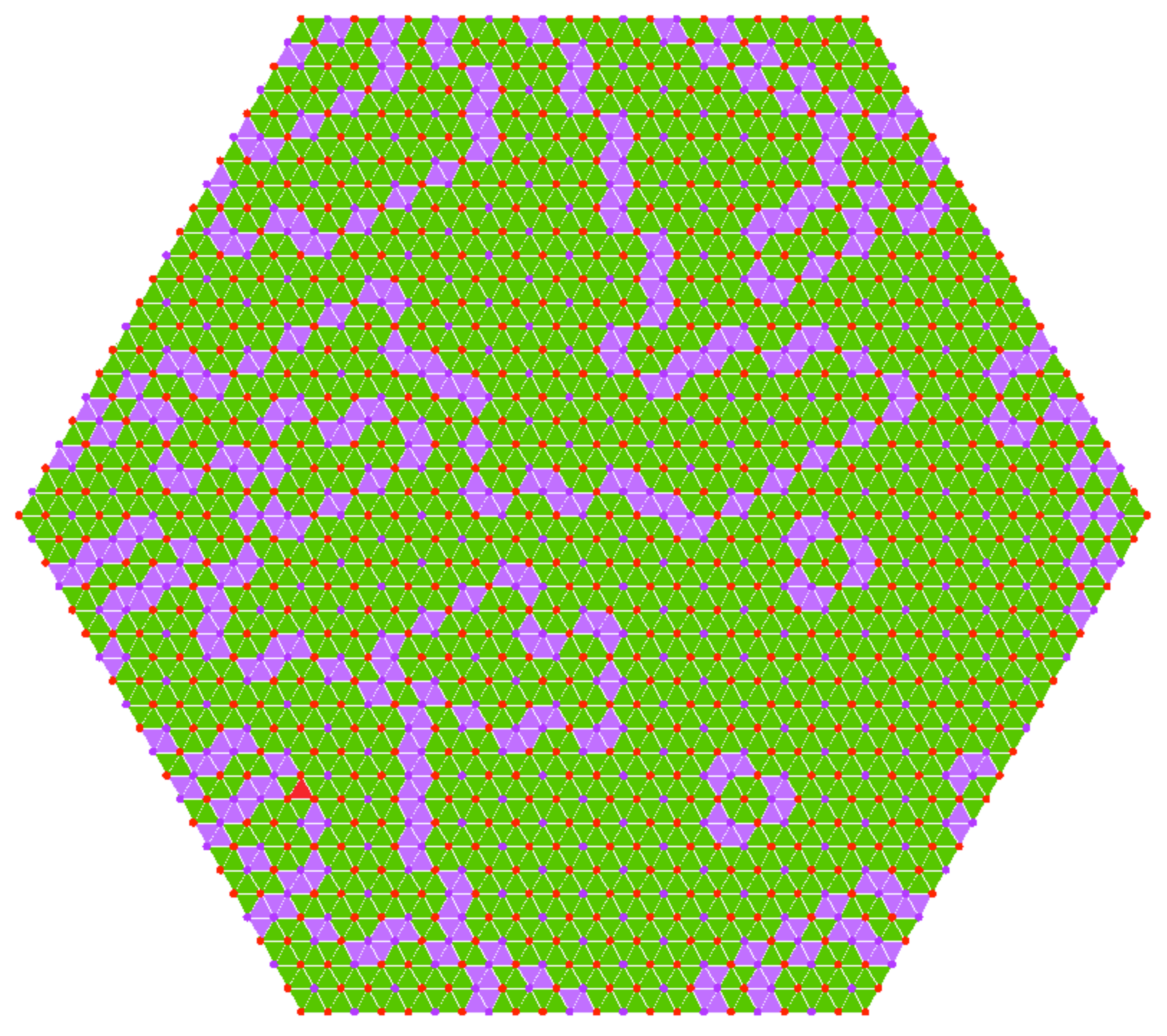}
\end{center}
\caption{Snapshot of a particular magnetic configuration in an hexagonal
cluster of 1387 spins, at the first sub-step, of field $H=0.81$~T, at
$T=4$~K and $10^4$ mcf, as depicted with a red line in Figure~\ref{mcfvar}.
Notice the ordered substrate green zone, with $M=M_s/3$, superimposed by
interlinked domain walls, formed mostly by connected triangles in violet
color. These domain walls contribute to a reduction of the $M_s/3$ value
of the green zone to $M\approx 0.22M_s$, in this example.}
\label{snap0.81T_4K}
\end{figure}

Configurations at zero field are formed by disordered bicontinuous patterns
of small compact regions of violet and green phases, with total zero magnetization.
For a field close to $0.25$~T, domain walls, as seen in Figure~\ref{snap0.81T_4K},
stabilize their motion to quasi-static domain walls until the field reach
values of about 1.2~T.

\begin{figure}[t]
\begin{center}
\includegraphics[scale=0.29,angle=0]{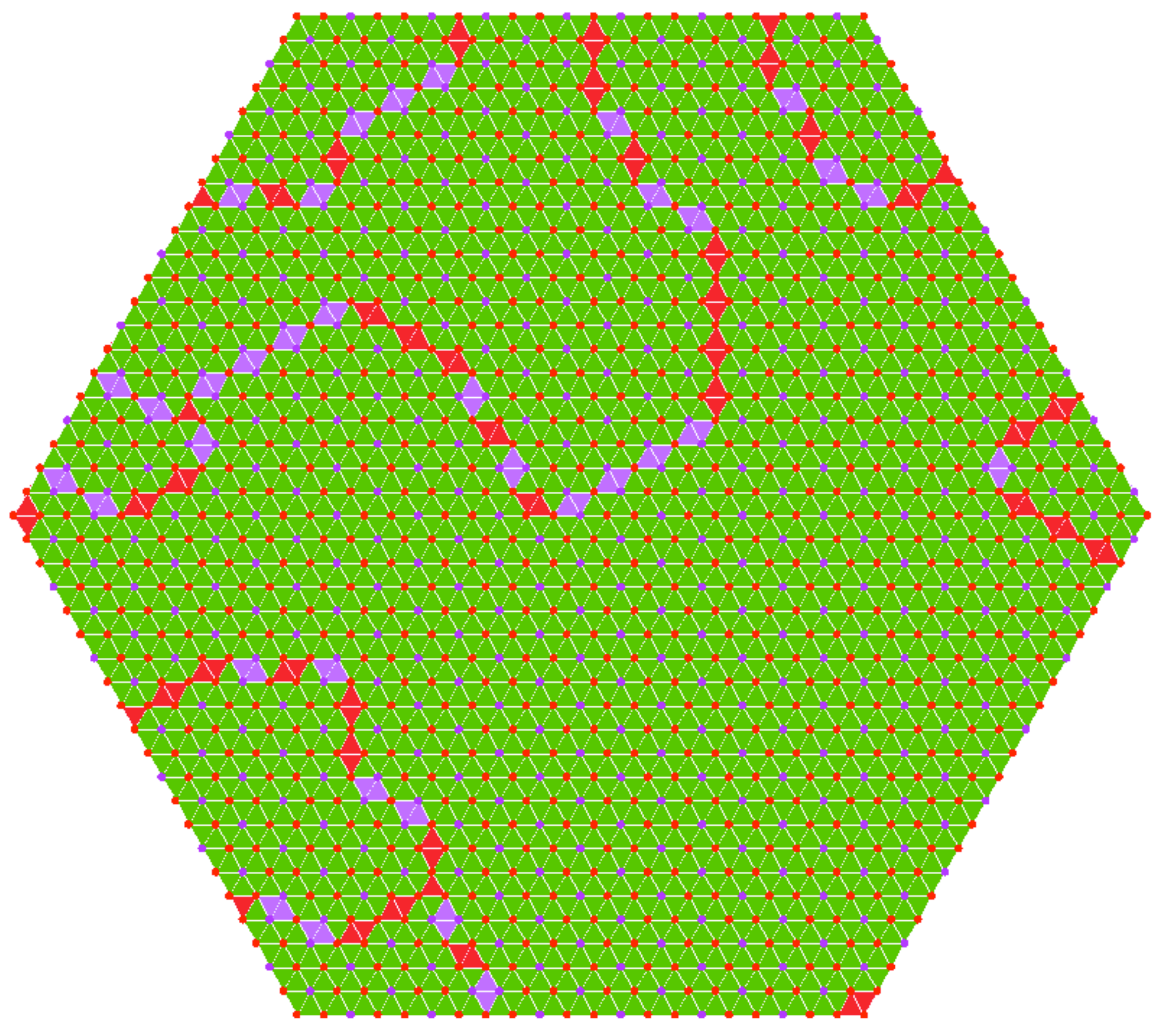}
\end{center}
\caption{Same parameters as in Figure \ref{snap0.81T_4K}, except at a higher
field $H=2.0$~T, corresponding to a magnetization value of $0.33M_s$ in this case.
Observe now the green extended zone superimposed by linked domain walls of mixed violet
and red colored triangles, in a proportion that contribute almost zero extra moment.
These linked domain walls, as those before, percolate throughout the system and are rather
stable across a finite window of the magnetic field: $1.2$~T~$<H<2.4$~T.}
\label{snap_2.0T_4K}
\end{figure}

Moving the field across the boundary, $H=1.2$~T, produces a rapid rearrangement
of the linked domain walls positions, by changing simultaneously their internal
configuration. They are now formed by mixed violet and red triangles, as shown
in Figure \ref{snap_2.0T_4K}. They rate together to zero extra magnetic moment,
sometimes oscilating above and below, depending slightly on the simulation
parameters. The magnetic configuration in Figure \ref{snap_2.0T_4K} averages
to zero, leading to $M=M_s/3$, the value of the green ordered zone (formed by
the hexagonal unit cells).

\begin{figure}[bt]
\begin{center}
\includegraphics[scale=0.29,angle=0]{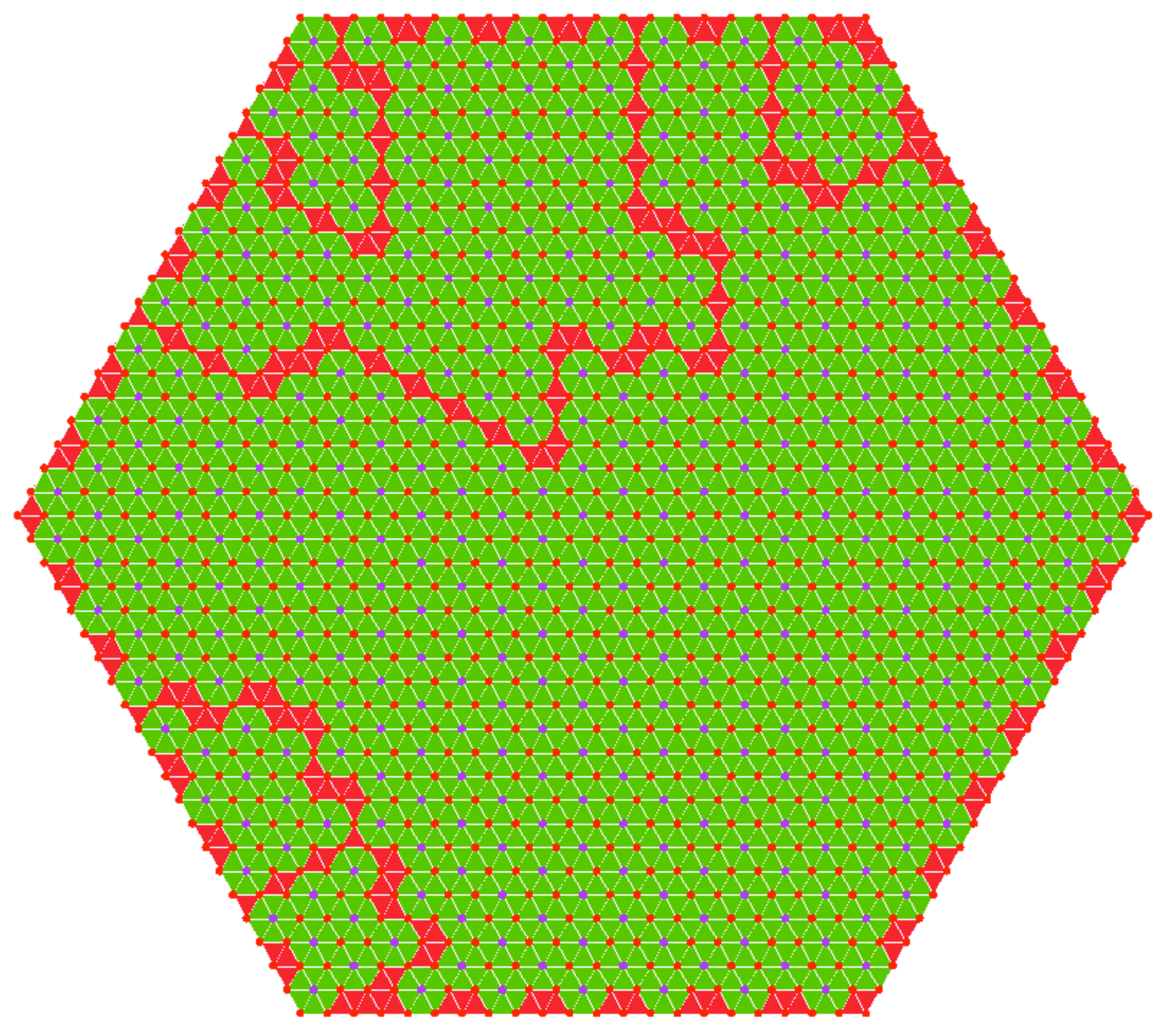}
\end{center}
\caption{Snapshot of the third sub-plateau, at a field $H=2.98$~T. Other
parameters as in Figures  \ref{snap0.81T_4K} and \ref{snap_2.0T_4K}.
The magnetic configuration is now formed by the green extended region
of $M=M_s/3$ superimposed by domain walls in red color, giving a total
magnetization of $0.41M_s$, in this case. Notice also some red seeds
distributed at the borders of the cluster, like in pre-nucleation
phases of irreversible studies. Nucleation really starts beyond the
critical field, $H_c=3.6$~T.} \label{snap_2.98T_4K}
\end{figure}

Finally, for fields beyond the other boundary, $H=2.4$~T, the situation is
quite different. Now the linked domain walls are formed mostly by red
colored triangles (Figure 9) with a tendency to coalesce in small groups, or seeds,
at the borders of the simulated clusters. This interesting feature of red seeds at the
boundaries, as seen in Figure~\ref{snap_2.98T_4K}, is a peculiar phenomena very similar
to pre-nucleation phases, as found in irreversible statistical mechanics studies.
Proper nucleation really starts beyond the critical field, $H_c=3.6$~T, in this way:
first coalescence of small red seeds at the borders and then myriads of red nuclei
forming and growing, taking account of the whole cluster, in the bulk, until saturation.
The nucleation process above $H_c$ is a very rapid, disordered one, destroying
the linked domain walls structures seen in the former examples. Perfect alignment
is only achieved at the end, for fields higher than 4 T.

All these non-equilibrium phenomena just described were obtained for a particular
sweep rate, of 10$^4$ mcf. We tested various sweep rates, observing similar behavior
for many of them. But, none of these metastable interlinked domain walls were present,
however, when we go beyond a sweep rate of 10$^9$ mcf. In that case, the disordered
initial phase goes immediately to the 1/3 plateau, and beyond $H_c=3.6$~T, to
saturation. Such 1/3 plateau is represented solely by the green zone, without any
further domain wall, thus of long-range nature. Therefore, we believe our
characterization of such metastable states, sweep-rate dependent, must be related
to the multi-step phases of Ca$_3$Co$_2$O$_6$, found experimentally at low $T$. 

\section{Triangular clusters}

 In this section, we show
how the 1/3 plateau of the Ising model of previous sections shares interesting
similarities with the simplest small clusters of triangular geometry. A generic
spin Hamiltonian for these clusters is the following Heisenberg Hamiltonian:
\begin{equation}
{\cal H} = \frac{1}{2}\sum_{ i,j}^{N} J_{ij}{{\bf S}_i}\!\cdot\!{{\bf S}_j}
-\delta \sum_{i}^{N}\left({{\bf S}_i}\!\cdot\!{\hat{\bf n}}\right)^2
- g{\mu_{\rm B}}\sum_{i}^{N} {{\bf S}_i}\!\cdot\!{\bf H}
\label{numeric}
\end{equation}

\begin{figure}[t]
\begin{center}
\includegraphics[scale=0.43,angle=0]{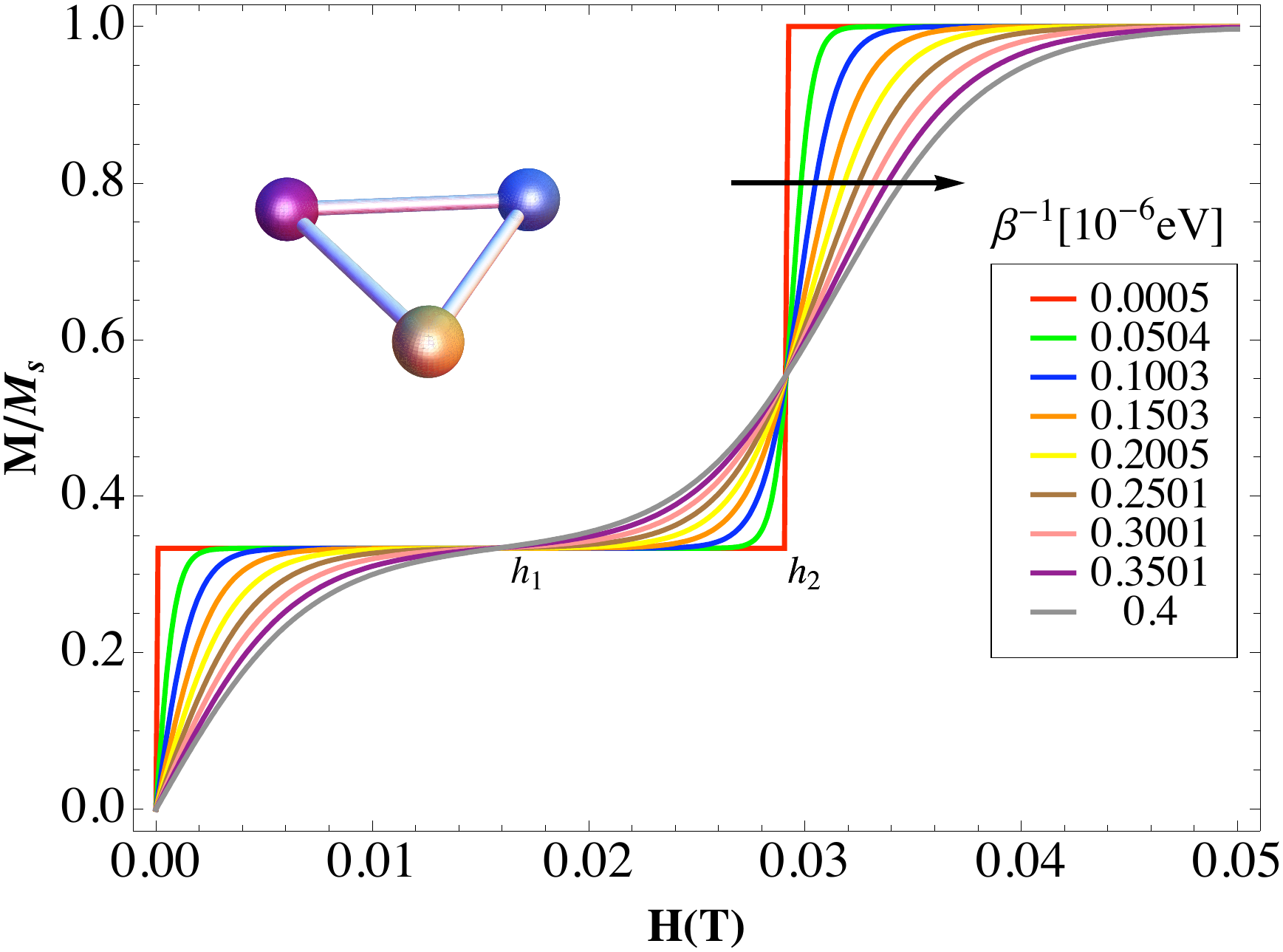}
\end{center}
\caption{Exact magnetization against a magnetic field for the
Heisenberg triangular molecule ($N=3$), and for different temperatures,
${\beta}^{-1}=k_{\rm B} T$, as shown in the legend. We see a tendency
to form a 1/3 plateau. Other parameters used are: $J=2.25$ $\mu$eV,
$\delta=1$ $\mu$eV, and $S=1/2$.}
\label{Triangulo}
\end{figure}

\noindent
Results for a triangular molecule are shown in Figure \ref{Triangulo}.
 $M$ (the thermal average of the total spin z-component) is zero only at zero field, and it increases rapidly   to the 1/3
plateau for any finite $H$, crossing through $H=h_1$ until $H=h_2$,
where it rapidly increases again to saturation. Such a transition is first-order at $T=0$ only.
Increasing the temperature produces a smooth growing up of $M$ due to thermal
fluctuations.  Notice that larger thermal fluctuations
accelerate this trend above $h_1$ and retard it above $h_2$, as seen from Figure
\ref{Triangulo}.

\begin{figure}[t]
\begin{center}
\includegraphics[scale=0.59,angle=0]{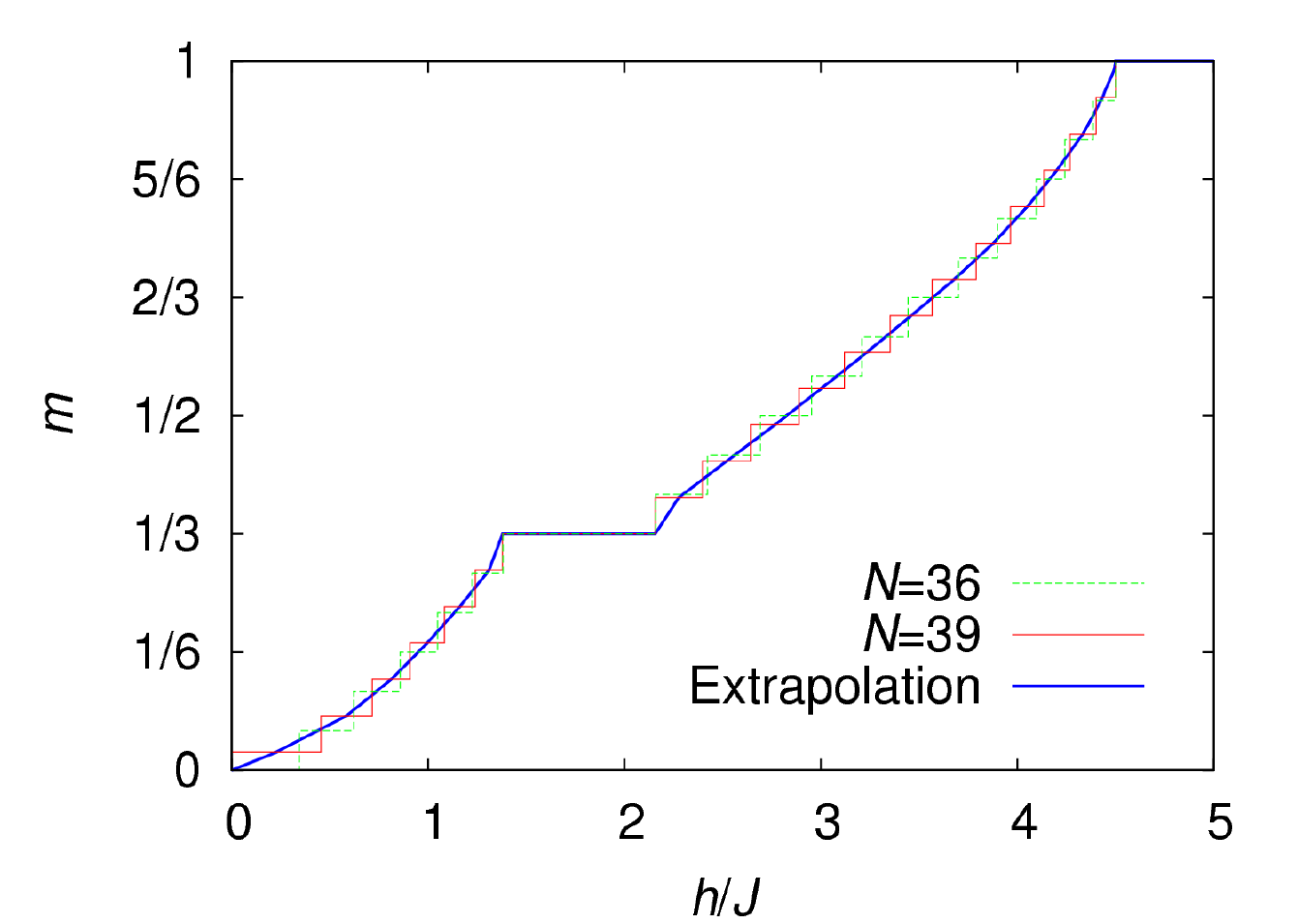}
\end{center}
\caption{Exact magnetization against a magnetic field for Heisenberg triangular
clusters at $T=0$, $\delta=0$, and $S=1/2$. The green and red lines are for
$N=36$ and 39 sites, while the bold blue line is an extrapolation to the
thermodynamic limit (see text). Notice the 1/3 plateau of finite width $\Delta H$ \cite{honeckerPC}.}
\label{exact_plateau}
\end{figure}

Interesting to note is that such a trend in Figure \ref{Triangulo}
resembles the MC solution of the Ising model in the triangular lattice of our
previous sections. Although it is not the same behavior, a tendency to form a 1/3
plateau is without any doubt. No further sub-plateaus can possibly be observed in a
triangular molecule because of lack of phase space. Domain walls in the Ising
dynamics are minimally formed by triangles, as seen before.

Further supporting arguments come from  numerical exact solutions for (larger) finite
clusters. In Figure \ref{exact_plateau} we see, for example, numerical results
for the magnetization of the pure Heisenberg model in the triangular lattice, in
the absence of anisotropy, at $T=0$, for $N=36$ and 39 sites \cite{honeckerPC}. The
formation of a clear 1/3 plateau of finite width, $\Delta H$, is seen from the graphic.
Such results have been taken in the literature \cite{bernu,honecker,honecker2,richter}
as the evidence of a definite plateau in the thermodynamic limit: the bold line
in Figure \ref{exact_plateau} was drawn by connecting the midpoints of
the finite-size steps for $N=39$ sites, except at the boundaries of the 1/3
plateau. A good estimate for the lower and upper fields of the boundaries
of the exact 1/3 plateau in the triangular lattice were obtained
comparing spin-wave and exact diagonalization results. They are given
by $H_\ell=1.378J$ and $H_u=2.155J$, respectively. See more recent
results in \cite{richter}.

To better understand our point, we should recall that both models, Ising and
Heisenberg, show a similar behavior -- with a 1/3 plateau -- because the magnetic
field already breaks the symmetry from $SU(2)$ to $U(1)$, thus making the strong
quantum fluctuations inoperant in the isotropic Heisenberg triangular case and,
consequently, the $M=1/3$ state becomes a classical, collinear state, with
long-range order.

 \section{Conclusion and final remarks}

In summary, we did an extensive investigation of the triangular lattice using
Monte Carlo simulation techniques. We studied this lattice using a 2D Ising model for
large magnetic moments coupled antiferromagnetically. We were motivated by
multiple steps observed in the magnetization of the spin-chain Ca$_3$Co$_2$O$_6$
compound. We found a very good agreement between our results and the experimental
situation \cite{MaignanPRB702004}. We demonstrated that the observed plateaus of the
ferrimagnetic phase, below $T \leq 10$~K, consist of metastable states and for this
reason are time-dependent.
In the experimental results such dependence was observed in the sweeping rate of the
magnetic field, and in our case this dependence comes from the number of Monte Carlo
flips (mcf). In fact, a close parallelism is found between these two approaches, as
seen from our Figures \ref{Tvar}, \ref{mcfvar}, \ref{T4K_5K}, and Figures 3, 4, 5 from
Ref. \cite{MaignanPRB702004}. Besides such agreement, we also found that in the limit
of very large number of mcf, namely, a very slow rate of the field, the three sub-steps
converge into one plateau, at $M/M_s=1/3$, like in the experiments, thus giving a thermal decay of such
metastable states into the equilibrium, final 1/3 plateau, which has long-range hexagonal
order (described by the green ordered zone of the snapshots, with an hexagon as unit cell).

Further analysis using snapshots of some specific configurations revealed the presence
of linked (mobile) domain-walls that give origin to the splitting of the 1/3 plateau
below $T\leq 10$~K, as explained in the text. Such a result disproves the possibility
of quantum tunneling as generating the multiple steps, at least for temperatures above
4~K, and for the reported sweep rates. Based on our results, we believe that at lower
temperatures an arrested configuration may provide a slower dynamics of the domain-walls,
that would explain a $T$-independent time-decay process below 4~K, by critical slowing down,
an aspect that certainly must still be checked down.

Our solution has been checked by further considering the intrinsic dipolar interactions,
finding no relevant contribution from such couplings. We therefore discarded them as possible
contribution of hysteresis phenomena which are strongly present at the ferrimagnetic region,
below 10~K, in the experiments \cite{MaignanPRB702004}. We believe that other ingredients
and methods, like 3D simulations,  must be considered to deal with such irreversible aspects, but this is beyond
our model of the magnetization plateaus.

Another point that we wanted to emphasize was linked to exact solutions of the
Heisenberg model in the triangular lattice for finite clusters. We did a comparison
with our previous numerical Ising solution and found supporting arguments with
the presence of the 1/3 plateau in the isotropic Heisenberg case at $T=0$.
Both models share the 1/3 plateau, although for different reasons. In the
quantum-mechanical case the applied magnetic field strengthens  the frustration
effects in the triangular lattice up to a point of having sufficient overlap with
the classical, collinear, three-fold degenerate, 1/3 plateau \cite{honecker2}. The point here is that
the presence of the 1/3 plateau in the isotropic quantum limit does not require a
strong uniaxial anisotropy. Although we did not a thermal study of the exact solution
of the Heisenberg model, a feature hard to be done, we think that at low temperatures
the 1/3 plateau would survive to thermal fluctuations in order to be observed numerically.
Such a comparison, therefore, strongly supports our thermal, classical study of the
spin-chain compound, Ca$_3$Co$_2$O$_6$.

While this paper was considered for publication we became aware of another work, by
Kudasov {\it et al.} \cite{KudasovPRB2008}, on the same problem, namely, a study of
the dynamics of magnetization in  frustrated spin-chain system Ca$_3$Co$_2$O$_6$
using a 2D model. Although they present similar results for the magnetization steps,
they are complementary to ours in some respects.

\acknowledgements

This work was partially supported by Fondecyt grant number 1070224, CONICYT Ph.D. Fellowship
Millennium Science Initiative under Project P06-022-F and also supported  by CNPq (Brasil).
One of us (GM) would like to acknowledge the
Universidad T\'ecnica Federico Santa Mar\'{\i}a, in Valpara\'{\i}so, Chile,
for a sabbatical stay. We also thank
Dr. Andreas Honecker for sharing his data on the triangular lattice.

\end{document}